\pdfoutput=1
\documentclass[aps,prb,preprint,floatfix,groupedaddress,amsmath,amssymb,amsfonts]{revtex4-1}
\usepackage{mathptmx}
\usepackage[scaled=0.9]{helvet}
\usepackage[utf8]{inputenc}
\usepackage{graphicx}
\usepackage[dvipsnames]{xcolor}
\usepackage{soul} % provides highlighting
\usepackage{textcomp} % provides \textmu µ
\usepackage[pdftex]{hyperref}
\hypersetup{pdftitle={Individual electron and hole localization in submonolayer InN quantum sheets embedded in GaN}, colorlinks, citecolor=blue}
\usepackage[
,textwidth=17.5cm
,textheight=23.5cm
,verbose
,pdftex
]{geometry}

\begin{document}

\title{Individual electron and hole localization in submonolayer InN quantum sheets embedded in GaN}
\author{F. Feix} \email[Electronic mail: ]{feix@pdi-berlin.de}
\author{T. Flissikowski}
\author{C. Ch\`{e}ze}
\author{R. Calarco}
\author{H. T. Grahn}
\author{O. Brandt}
\affiliation{Paul-Drude-Institut f{\"u}r Festk{\"o}rperelektronik, Leibniz-Institut im Forschungsverbund Berlin e.~V.,
Hausvogteiplatz 5--7, 10117 Berlin, Germany}

\begin{abstract}
We investigate sub-monolayer InN quantum sheets embedded in GaN(0001) by temperature-dependent photoluminescence spectroscopy under both continuous-wave and pulsed excitation. Both the peak energy and the linewidth of the emission band associated with the quantum sheets exhibit an anomalous dependence on temperature indicative of carrier localization. Photoluminescence transients reveal a power law decay at low temperatures reflecting that the recombining electrons and holes occupy spatially separate, individual potential minima reminiscent of conventional (In,Ga)N(0001) quantum wells exhibiting the characteristic disorder of a random alloy. At elevated temperatures, carrier delocalization sets in and is accompanied by a thermally activated quenching of the emission. We ascribe the strong nonradiative recombination to extended states in the GaN barriers and confirm our assumption by a simple rate-equation model.
\end{abstract}

\maketitle

The ternary alloy (In,Ga)N enables the realization of efficient light emitters used for solid-state lighting,\cite{Narukawa2010} display technologies,\cite{Pust2015} and diode lasers.\cite{Nakamura1996,Izumi2015} The high efficiency is believed to be linked to carrier localization due to the inevitable compositional fluctuations occurring on an atomic scale in the random alloy (In,Ga)N.\cite{Brandt2006} This phenomenon, however, also results in a severe inhomogeneous broadening of electronic transitions and a strongly retarded recombination dynamics.\cite{Morel2003,Brosseau2010,Watson-Parris2011a} Both of these effects are detrimental for laser applications.\cite{Chow1997,Witzigmann2006a}

Digital alloys composed of an InN/GaN short-period superlattice (SPSL)\cite{Yoshikawa2007,Dimakis2008} are envisioned to eliminate alloy disorder and the resulting localization phenomena. Thus, these structures are expected to exhibit a much reduced inhomogeneous broadening and an enhanced radiative recombination rate. However, a recent microscopic investigation of InN/GaN SPSLs demonstrated that the nominal InN monolayers (MLs) in the SPSL have a coverage of only 0.33\,MLs.\cite{Suski2014} This finding raises the question whether the InN quantum sheets (QSs) consist of two-dimensional InN islands of nanometer size or of a single monolayer of disordered In$_{0.33}$Ga$_{0.67}$N.

A third possibility arises from the existence of an In adatom-induced $\left(\sqrt{3}\times\sqrt{3}\right)$\text{R}30$^\circ$ surface reconstruction on GaN(0001) that is expected to be formed by a third of a ML of In,\cite{Chen2000,Waltereit2002} i.\,e., the same coverage as observed by \citet{Suski2014} This agreement may be coincidental, but it may also suggest that, under the usual growth conditions, the In coverage in the first ML of InN growth is restricted to the one that constitutes the energetically favorable surface phase, namely, the $\left(\sqrt{3}\times\sqrt{3}\right)$\text{R}30$^\circ$-In adsorbate structure.\cite{*[{Note that a reconstruction with this symmetry has also been observed for (In,Ga)N by  }] [{ as well as for pure InN by }] Friedrich2012,*Himmerlich2009}
Since this surface phase is self-limiting in thickness to a single ML and is laterally ordered, it may provide a template for the insertion of ordered InGa$_2$N$_3$ QSs in GaN.\cite{Gorczyca2015a}

In this Letter, we study samples intentionally fabricated under conditions ensuring that the InN QSs are formed by this In adlayer and not by actual InN growth.
We employ temperature-dependent photoluminescence (PL) spectroscopy under both continuous-wave (cw) and pulsed excitation to explore the electronic properties of these structures, which are found to be essentially indistinguishable from those reported in the literature for nominally full InN MLs embedded in GaN. Moreover, our experiments demonstrate unambiguously that the sub-ML InN QSs act electronically as two-dimensional random alloys rather than ordered InGa$_2$N$_3$ or InN ML islands.

We employed plasma-assisted molecular beam epitaxy to produce samples with periodically inserted InN QSs with the same nominal coverage, but a different vertical separation. These heterostructures were deposited onto GaN(0001) templates at a substrate temperature of 550\,$^\circ$C,
i.\,e., significantly above the decomposition temperature of InN(0001).\cite{Gallinat2007} Combining high-resolution electron microscopy and x-ray diffracto\-metry, samples I and II were found to consist of ten individual QSs restricted in width to a single ML with an In coverage of 0.29\,MLs and separated by 6 and 50\,MLs of GaN, respectively.
For cw-PL spectroscopy, the samples were excited by a He-Cd laser ($\lambda_\text{L} = 325$\,nm) with an excitation power density of 100\,Wcm$^{-2}$.
For the time-resolved PL (TRPL) experiments, a frequency-doubled, femtosecond Ti:sapphire laser ($\lambda_\text{L} = 349$\,nm) was used to create pulses with an energy fluence per pulse of 50\,\textmu J\,cm$^{-2}$ and a repetition rate of 420\,kHz. We employed time-correlated single-photon counting providing a dynamic range in detection of more than four orders of magnitude. For both the cw and TRPL experiments, the samples were mounted in a He-flow cryostat allowing a continuous variation of temperature between 10 and 300\,K.

\begin{figure}[b]
\includegraphics[width=0.5\columnwidth]{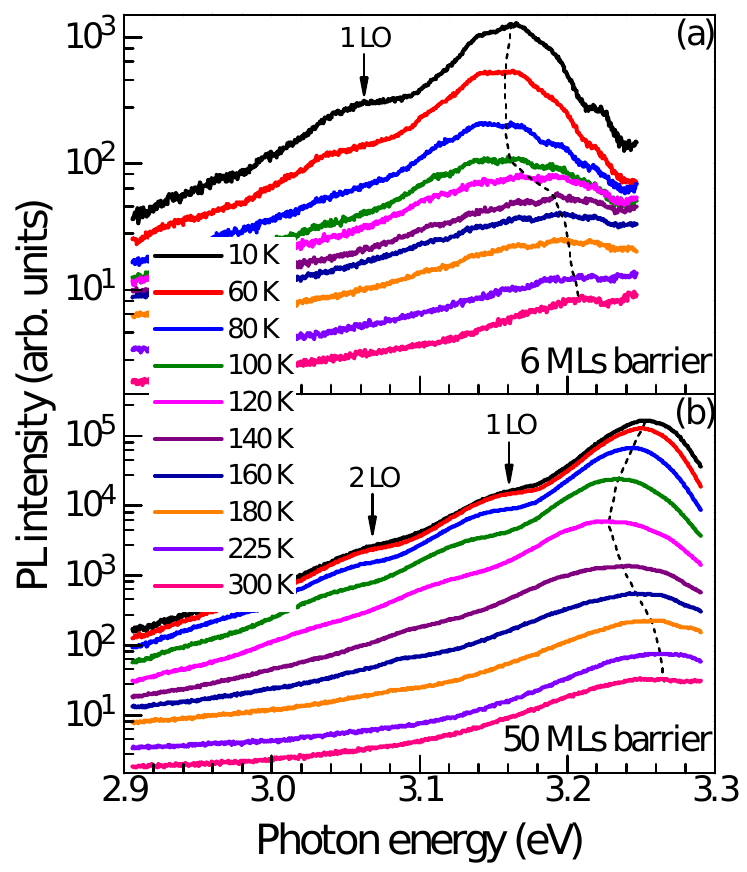}
\caption{Temperature-dependent cw-PL spectra of (a) sample~I (6\,MLs barrier) and (b) sample~II (50\,MLs barrier). The dashed lines indicate the peak position of the QS luminescence and the arrows the first as well as the second order longitudinal optical (LO) phonon replica.}
\label{fig:spectra}
\end{figure}
Figures \ref{fig:spectra}(a) and \ref{fig:spectra}(b) show temperature-dependent cw-PL spectra of samples I and II, respectively, on a semilogarithmic scale. The PL band peaks at around 3.16\,eV for sample~I and at 3.25\,eV for sample~II. These energies for our samples with sub-ML coverage are close to those reported for nominally complete InN MLs (see Ref.~\onlinecite{Staszczak2013} as well as Ref.~\onlinecite{Suski2014} and references therein), suggesting that these latter samples do not have full ML coverage either. They furthermore agree very well with the values calculated in Refs.~\onlinecite{Gorczyca2013,Suski2014,Gorczyca2015a} for SPSLs with an In coverage of 0.33\,MLs. In particular, we observe the predicted redshift in transition energy by \citet{Suski2014} for thin barriers (our sample I) as a result of the electronic coupling between the electron states in the QSs. Note, however, that the peak energies do not follow the behavior of the monotonically decreasing band gap of (In,Ga)N with increasing temperature as indicated by the dashed lines in Fig.~\ref{fig:spectra}. Note further that the line width of the PL band of 60\,meV is not expected for the emission from an ordered layer, but is rather indicative of disorder. Even larger line widths have been reported for InN/GaN samples with nominally complete InN MLs.\cite{Staszczak2013}

\begin{figure}[b]
\includegraphics[width=0.4\columnwidth]{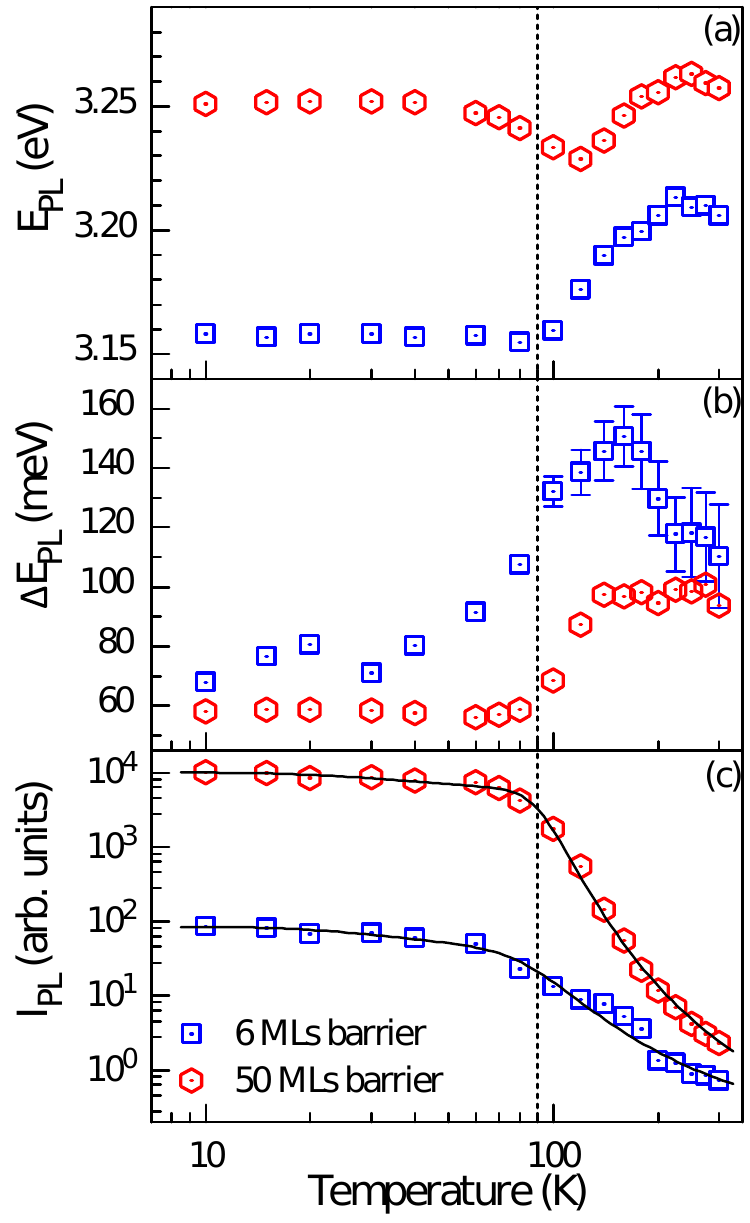}
\caption{\label{fig:analysis}(a) Temperature-dependent peak energy $E_{\text{PL}}$, (b) full width at half maximum $\Delta E_{\text{PL}}$ and (c) integrated PL intensity $I_{\text{PL}}$ of the QS PL band of sample~I with 6\,MLs barrier (blue squares) and sample~II with 50\,MLs barrier (red hexagons). The dashed line indicates the threshold temperature at which all three quantities change. The solid lines in (c) represent Arrhenius fits with activation energies of 42 and 89\,meV for samples~I and II, respectively.}
\end{figure}

Figure \ref{fig:analysis} shows the results of a line shape analysis of the PL spectra shown in Fig.~\ref{fig:spectra}. At low temperatures, the emission bands of samples~I and~II differ significantly not only in the peak PL energy $E_{\text{PL}}$, but also in the integrated intensity $I_{\text{PL}}$. This difference is partly due to the fact that sample~II absorbs about 79\% of the incident photons within the total thickness of 132\,nm of the 10-period InN/GaN superlattice as compared to 19\% for sample~I with its total thickness of only 18\,nm. However, this effect accounts only for a factor of four. In fact, the integrated intensity obtained with quasi-resonant excitation ($\lambda_\text{L} = 363$\,nm), for which the total absorbance of the ten QSs is expected to be very similar for samples I and II, still differs by a factor of 50 between the two samples (not shown). A stronger influence of surface-induced electric fields for sample~I, causing carriers to escape from the QSs and to subsequently recombine at the surface nonradiatively, can be excluded as well, since samples identical to sample~I except for a 40\,nm thick cap layer exhibit a comparable PL intensity.

We believe that the large difference in PL intensity for samples I and II originates from the electronic coupling between the QSs in sample~I. The redshift of $E_{\text{PL}}$ by 90\,meV at 10\,K [cf.~Fig.~\ref{fig:analysis}(a)] indicates that this coupling is strong, which in turn means that the electron state is basically delocalized over the entire SPSL. This fact also increases the probability of electrons to reside in the barriers and, thus, as we will see below, to suffer from nonradiative recombination. Furthermore, for an SPSL with a finite number of QSs, the large electrostatic fields within the QSs will not be perfectly balanced by opposing fields in the barriers, i.\,e., the heterostructure will exhibit a potential staircase which in turn will result in a vertical electron-hole separation.

With increasing temperature, $E_{\text{PL}}$ redshifts by about 20\,meV for sample~II with the thicker barriers, before the band blueshifts by 30\,meV. In contrast, $E_{\text{PL}}$ blueshifts monotonically by about 50\,meV for sample~I with the thinner barriers. Simultaneously with the change in transition energy, the full width at half maximum $\Delta E_{\text{PL}}$ increases by about 40\,meV (70\,meV), and $I_{\text{PL}}$ decreases by 2 (3.5) orders of magnitude for sample~I (sample~II). These simultaneous changes of the three main characteristics of the emission band, highlighted by the dashed line in Fig.~\ref{fig:analysis}, suggest that they have a common origin.

The evolution of the transition energy with temperature observed for sample~II [cf.~Fig.~\ref{fig:analysis}(a)] clearly resembles the well-known \textsf{S} shape commonly observed for random (In,Ga)N alloys.\cite{Eliseev1997} This behavior is most frequently ascribed to carrier localization at low temperatures.\cite{*[{See }] [{ and references therein.}] Hammersley2012} At elevated temperatures, carriers are able to relax from shallow to deep states (causing a decrease in the transition energy, such as observed here for $E_{\text{PL}}$ around 90\,K), while, at even higher temperatures, they are thermally activated to higher energy states (resulting in an increase in $E_{\text{PL}}$) and become mobile within the band of interacting localized states. The latter phenomenon is usually accompanied by an abrupt broadening of the emission band, as also observed here [cf.~Fig.~\ref{fig:analysis}(b)]. Furthermore, the delocalization of carriers that occurs at this point frequently results in an onset of nonradiative recombination, which manifests itself by an abrupt reduction of the luminous efficiency with increasing temperature.\cite{Dhar2002} Indeed, the integrated intensities of the PL bands of samples I and II start to decrease at 90\,K and are described well by an Arrhenius law with activation energies of $(42 \pm 10)$ and $(89 \pm 15)$\,meV for sample I and II, respectively [cf.~Fig.~\ref{fig:analysis}(c)].

\begin{figure}[b]
\includegraphics[width=0.8\columnwidth]{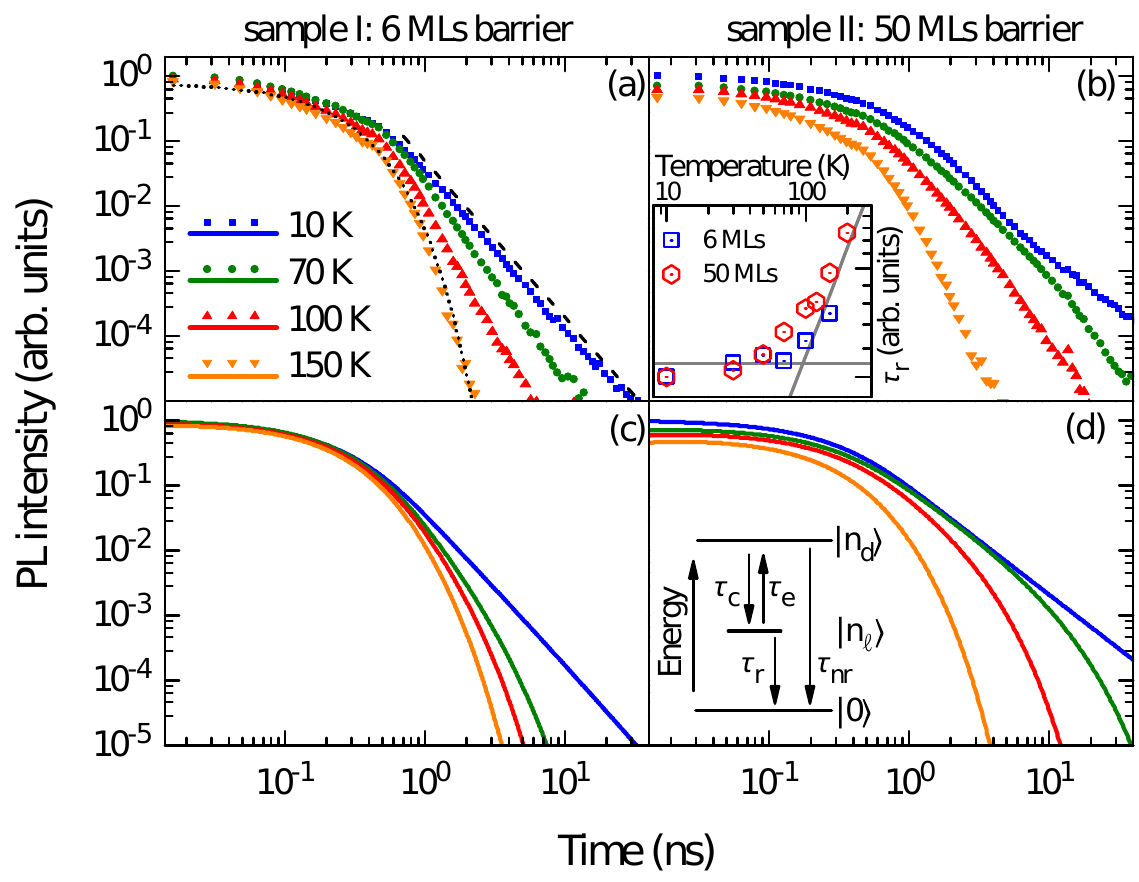}
\caption{\label{fig:trpl}Experimental PL transients of (a) sample~I and (b) sample~II at various temperatures as indicated in (a). The dashed and dotted lines in (a) show a comparison with a $T^{-2.4}$ power law and a fit with a single exponential, respectively. The inset in (b) displays the temperature dependence of the inverse peak intensity $I_\text{max}$ of the transient, which is proportional to the radiative lifetime $\tau_r$. The lines are a guide to the eye. The corresponding simulated PL transients are shown in (c) and (d). The simulations are based on the rate-equation system schematically depicted as the inset in (d).}
\end{figure}

An alternative explanation for the simultaneous blueshift of the emission band and a quenching of its intensity as observed for (In,Ga)N/GaN(0001) quantum wells (QWs) has been recently proposed by \citet{Langer2014a} Their model relies on the exponential variation of the radiative lifetime with transition energy due to the strong piezoelectric fields within the QWs. An activated nonradiative recombination preferentially quenches the longer living transitions, resulting in an effective blueshift of the emission band. Within this interpretation, the activation energy deduced from the data in Fig.~\ref{fig:analysis}(c) would be related to the activation of nonradiative centers and not to the localization energy of carriers.

In order to distinguish between these two different interpretations of our data, we performed temperature-dependent TRPL experiments. We observe similar transients regardless of the specific emission energy. Figures~\ref{fig:trpl}(a) and \ref{fig:trpl}(b) show the decay of the PL intensity at the peak energy of the emission band of samples I and II, respectively, in a double logarithmic representation. The recombination dynamics observed is qualitatively similar for both samples and is characterized by the following three major properties. First, the peak PL intensity of the transient just after the laser pulse $I_\text{max}$ is almost constant up to 50--70\,K and decrease thereafter with increasing temperature. Second, the decay at 10~K is very slow for both samples and actually follows a power law [cf.~Fig.~\ref{fig:trpl}(a)--(b)]. Third, the decay accelerates with increasing temperature and gradually approaches a monoexponential dependence at 150\,K [see the corresponding fit in Fig.~\ref{fig:trpl}(a)]. In the following, we will interpret these three observations.

(i) Since $I_\text{max}$ is proportional to the inverse radiative lifetime,\cite{Brandt1996} it follows that the radiative lifetime is almost constant up to about 50--70\,K as shown in the inset of Fig.~\ref{fig:trpl}(b). A temperature-independent radiative lifetime is a fingerprint for transitions arising from zero-dimensional states.\cite{Waltereit2001} At higher temperatures, the radiative lifetime approaches the linear increase expected for radiative transitions in two-dimensional systems such as QWs and QSs [see the inset of Fig.~\ref{fig:trpl}(b)].\cite{Waltereit2001} These results thus support the interpretation of the data shown in Fig.~\ref{fig:analysis} in terms of carrier localization at low temperatures followed by delocalization at 70--90\,K.

(ii) The power law decay kinetics observed for both samples at 10\,K furthermore demonstrates that the recombination is not excitonic, but takes place between individually localized electrons and holes with varying spatial separation.\cite{Morel2003,Brosseau2010,Sabelfeld2015} In fact, theoretical studies of (In,Ga)N quantum wells (QWs) have predicted that already the inevitable compositional fluctuations of a random (In,Ga)N alloy induce strong hole localization, whereas electrons are localized rather by fluctuations of the QW width.\cite{Watson-Parris2011a,Schulz2015b} The initial decay is then due to the recombination of electron-hole pairs with minimum spatial separation, but, as the decay proceeds, recombination can only take place between remote pairs of progressively increasing distance and thus slows down.\cite{Sabelfeld2015}

(iii) Together with the acceleration of the decay at elevated temperatures, the time-integrated intensity of the transients decreases significantly, suggesting that both observations are linked by a nonradiative channel made accessible by the delocalization of carriers. To test this hypothesis, let us consider a simple model for the recombination dynamics visualized by the level scheme in the inset in Fig.~\ref{fig:trpl}(d). We implicitly assume that the recombination is determined by holes populating either localized $\vert{n_\ell}\rangle$ or extended/delocalized states $\vert{n_d}\rangle$. These populations are coupled via the relaxation of holes from $\vert{n_d}\rangle$ to $\vert{n_\ell}\rangle$ with a time constant $\tau_c$ and by their thermally activated emission from $\vert{n_\ell}\rangle$ to $\vert{n_d}\rangle$ with a time constant $\tau_e$. To keep the model as simple as possible, we assume that recombination from localized states is purely radiative with a temperature-dependent lifetime $\tau_r(T)$ as observed experimentally, while the extended states are supposed to be dominated by a nonradiative process with a constant lifetime $\tau_{nr}$.

These considerations lead to the following coupled system of differential equations
\begin{align}
	\label{eq1}
    n_d(T)' & = -\frac{n_d}{\tau_c} - \frac{n_d}{\tau_{nr}} + \frac{n_\ell}{\tau_e} \exp\left(-\frac{E_b}{k_\text{B} T}\right)\\
    \label{eq2}
    n_\ell(T)' &= \frac{n_d}{\tau_c} - t^{a-1} \frac{n_\ell}{\tau_r(T)} - \frac{n_\ell}{\tau_e} \exp\left(-\frac{E_b}{k_\text{B} T}\right)
\end{align}

\noindent
with the activation energy $E_b$ for the emission of holes from localized to extended states. The prefactor $t^{a-1}$ of the radiative term $t^{a-1} n_\ell/\tau_r$ in Eq.~(\ref{eq2}) results in a stretched exponential decay, but approaches a power law decay for $a \rightarrow 0$.

Intensity transients simulated by Eqs.~(\ref{eq1}) and (\ref{eq2}) are shown in Figs.~\ref{fig:trpl}(c) and \ref{fig:trpl}(d) for sample~I and II, respectively. Evidently, the experimentally observed evolution from a power law to a monoexponential decay together with the simultaneous loss in intensity are well reproduced. Our understanding of this phenomenon (namely, that it is solely induced by delocalization) is thus confirmed. Quantitatively, we have assumed activation energies $E_b$ of 16\,meV for sample~I and 27\,meV for sample~II to obtain a change with temperature in agreement with the experiments. Both of these values are a factor of about 3 smaller than those derived from our cw-PL experiments [cf.~Fig.~\ref{fig:analysis}(c)]. We ascribe this finding to the much higher (two orders of magnitude) excitation density in the TRPL experiments, which is known to significantly increase the actual carrier temperature with respect  to the temperature of the lattice.

In conclusion, the sub-ML InN QSs under investigation have been found to act electronically as two-dimensional random alloys similar to conventional (In,Ga)N QWs. Since the transition energies observed for our samples are essentially identical to those reported in the literature, we suggest that these nominally complete InN MLs are in fact also formed by the energetically favorable In adlayer on GaN(0001) with a maximum coverage of 0.33\,MLs but without detectable lateral ordering. Hence, the fabrication of InN/GaN digital alloys may be more difficult than envisioned.

We thank Tobias Schulz and Martin Albrecht (Leibniz-Institut f{\"u}r Kristallz{\"u}chtung, Berlin) for transmission electron microscopy and Carsten Pf{\"u}ller for a critical reading of the manuscript. Financial support of this work within the European Union Framework Programs Horizon-2020 under grant agreement No.\ 642574 (SPRInG) and FP7-NMP-2013-SMALL-7 under grant agreement No.\ 604416 (DEEPEN) is gratefully acknowledged.

\end{document}